\documentclass[aps,prb,reprint,superscriptaddress,floatfix]{revtex4-1}
\usepackage[utf8]{inputenc}

\usepackage{amssymb,amsmath,bm,graphicx}
\usepackage{ifpdf,color,mathrsfs}

\graphicspath{{./figures/}}

\definecolor{darkblue}{rgb}{0,0,.6}
\definecolor{darkgreen}{rgb}{0,0.5,0}

\ifpdf
\usepackage{epstopdf}
\usepackage[pdftex,unicode,
pdfstartview={FitH},pdfborder={0 0 0}]{hyperref}
\usepackage{hypcap}
\else
\usepackage[hypertex]{hyperref}
\fi
\hypersetup{
	bookmarksnumbered = true,
	colorlinks = true, linkcolor = darkblue,
	citecolor = darkblue, filecolor = darkblue,
	menucolor = darkblue, urlcolor = darkblue
}

\newcommand{\scalarproduct}[2]{\bigl(#1 \cdot #2 \bigr)}
\newcommand{\iu}{\mathrm{i}}	
\newcommand{\de}{\mathrm{d}}	
\newcommand{\ee}{\mathrm{e}}	

\DeclareMathOperator{\re}{Re}

\newcommand{\pdiff}[2]{\frac{\partial\,#1}{\partial #2}}

\newcommand{\ket}[1]{\ensuremath{\left| #1\right\rangle}}
\newcommand{\braket}[2]{\ensuremath{\left\langle #1 | #2\right\rangle}}

\begin{document}
\title{Adiabatic corrections for velocity-gauge simulations of electron dynamics in periodic potentials}

\author{Vladislav S.~Yakovlev}
\email[]{vladislav.yakovlev@mpq.mpg.de}
\affiliation{Max-Planck-Institut f\"ur Quantenoptik, Hans-Kopfermann-Str. 1, 85748 Garching, Germany}
\affiliation{Ludwig-Maximilians-Universit\"at, Am Coulombwall~1, 85748 Garching, Germany}

\author{Michael S.~Wismer}
\affiliation{Max-Planck-Institut f\"ur Quantenoptik, Hans-Kopfermann-Str. 1, 85748 Garching, Germany}

\begin{abstract}
  We show how to significantly reduce the number of energy bands required to model the interaction of light with crystalline solids in the velocity gauge.
  We achieve this by deriving analytical corrections to the electric current density.
  These corrections depend only on band energies, the matrix elements of the momentum operator, and the macroscopic vector potential.
  Thus, the corrections can be evaluated independently from modeling the interaction with light.
  In addition to improving the convergence of velocity-gauge calculations, our analytical approach overcomes the long-standing problem of divergences in expressions for linear and nonlinear susceptibilities.
\end{abstract}

\maketitle

\section{Introduction}
The velocity and length gauges are two most frequent choices for a theoretical description of the interaction between electromagnetic radiation and matter.
Even though, in principle, all physical observables must be gauge-independent, the gauge choice is very important for numerical simulations because approximations and discretization errors violate the gauge invariance in its strict sense \cite{Scully_Quantum_Optics_chapter5}.
In atomic and molecular physics, where quantum systems are finite, both gauges are easily implemented. 
The situation is different for modeling the interaction of light with bulk crystals, where an infinite periodic lattice potential and the degenerate energy states present difficulties for length-gauge implementations~\cite{Aversa_PRB_1995,Virk_PRB_2007}.
These difficulties also pertain to the closely related approach of modeling quantum dynamics in the basis of Houston states~\cite{Houston_PR_1940}, also known as accelerated Bloch states~\cite{Krieger_PRB_1986}.
The fundamental problem here is that these methods require differentiation with respect to the crystal momentum, $\mathbf{k}$, while numerically evaluated Bloch states are, in general, not smooth functions of $\mathbf{k}$~\cite{Marzari_RMP_2012,Attaccalite_PRB_2013}.
The problem becomes particularly severe for modeling strong-field phenomena, where electrons traverse a significant part of the Brillouin zone during a laser cycle~\cite{Ghimire_NP_2011,Schubert_Nature_2014,Luu_Nature_2015,Wismer_PRL_2016}.
Solutions to these problems are known~\cite{Virk_PRB_2007,Gruening_PRB_2016}, but they either do not ensure the periodicity with respect to $\mathbf{k}$~\cite{Lindefelt_SST_2004} or require the evaluation of so-called covariant derivatives~\cite{Nunes_PRB_2001,Souza_PRB_2004,Virk_PRB_2007}.
Evaluating the covariant derivatives at each propagation step slows down computations.
In contrast, the velocity gauge does not require differentiation with respect to the crystal momentum, which often makes it advantageous for modeling light-driven electron dynamics~\cite{Yabana_PRB_2012,Goncharov_JCP_2013,Simonsen_JPB_2014,Wachter_PRL_2014,Yakovlev_SR_2015,Chizhova_PRB_2016}.
However, performing velocity-gauge simulations in a basis of Bloch states has a serious drawback: for convergence, they usually require many more energy bands than length-gauge or Houston-basis simulations~\cite{Wu_PRA_2015}.
In this paper, we propose a solution to this problem.
We do so by deriving analytical corrections to the polarization response evaluated with a relatively small number of bands (section \ref{sec:main_results}).
Although we derive our corrections in the limit of a weak external field, we show in section \ref{sec:example} that they work surprisingly well even for strong optical fields.

\section{Velocity-gauge description of light-solid interaction}
Let $\ket{n \mathbf{k}}$ denote a Bloch state with band index $n$ and crystal momentum $\mathbf{k}$.
The Bloch states are eigenstates of an unperturbed Hamiltonian $\hat{H}^{(0)}$.
That is, $\hat{H}^{(0)} \ket{n \mathbf{k}} = \epsilon_n(\mathbf{k}) \ket{n \mathbf{k}}$, where $\epsilon_n(\mathbf{k})$ is energy.
Even though $\hat{H}^{(0)}$ is a single-particle operator, it may represent an outcome of self-consistent many-body calculations.
We describe the interaction with a classical electromagnetic field by adding an interaction operator, $\hat{H}_{\mathrm{int}}$, to the Hamiltonian.
Unitary transformations allow one to write $\hat{H}_{\mathrm{int}}$ in different forms.
One of them is $\hat{H}_{\mathrm{int}} = e (\mathbf{A} \cdot \hat{\mathbf{p}} + \hat{\mathbf{p}} \cdot \mathbf{A}) / (2 m_0) + e^2 A^2 / m_0$.
Here, $\mathbf{A}$ is the vector potential of the electromagnetic field, $\hat{\mathbf{p}}$ is the momentum operator, $e>0$ is the elementary charge, and $m_0$ is the electron mass.
In SI units, the relation between $\mathbf{A}$ and the electric field is $\mathbf{E} = - \partial\mathbf{A}/ \partial t$.
In the following, we assume the dipole approximation, where the dependence of $\mathbf{A}$ on coordinate, $\mathbf{r}$, is neglected.
The term $e^2 A^2 / m_0$ can be eliminated by the unitary transformation $\ket{\psi} \to \exp[-\iu e^2 / (2 \hbar m_0) \int_{-\infty}^{t} A^2(t') \de t'] \ket{\psi}$.
The results is the velocity-gauge form of the Hamiltonian:
\begin{equation}
\label{eq:H_int}
\hat{H}(t) = \hat{H}^{(0)} + \hat{H}_{\mathrm{int}} =
  \hat{H}^{(0)} + \frac{e}{m_0} \mathbf{A}(t) \cdot \hat{\mathbf{p}}.
\end{equation}
In this equation, the external field preserves the spatial periodicity of the lattice potential (which may or may not be local).
Therefore, the Bloch theorem applies not only to $\hat{H}^{(0)}$, but also to $\hat{H}(t)$.
Consequently, transitions induced by the external field formally preserve the crystal momentum, which is one of the most important reasons why velocity-gauge simulations are so attractive for numerical calculations.

However, the convenience of preserving the crystal momentum  comes at a price.
This becomes apparent if we consider the interaction of a dielectric with a constant electric field ($\mathbf{E}=\text{const}$) that is sufficiently weak to neglect interband tunneling.
Even though, in the absence of free charge carriers, all physical observables are expected to take constant values, the velocity-gauge wave functions have a nontrivial dependence on time because the vector potential, $\mathbf{A}(t)=-\int_{-\infty}^{t} \mathbf{E}(t') \de t'$, makes the Hamiltonian time-dependent.
In this case, numerical approximations, such as truncating the basis, lead to a spurious divergence of the polarization response in the low-frequency limit~\cite{Aspnes_PRB_1972,Moss_PRB_1990,Sheik-Bahae_IEEE_JQE_1991,Sipe_PRB_1993,Aversa_PRB_1995}.
This is one of the underlying reasons why velocity-gauge calculations require many more bands than length-gauge or Houston-basis ones. 

Another, closely related, problem with the velocity gauge can be understood by considering a particular Bloch state exposed to a weak long half-cycle of the electric field.
Because the field performs only a half of the oscillation, the vector potential has a nonzero value at the end of the pulse.
In the adiabatic limit, the pulse transforms an initial state $\ket{n \mathbf{k}_0}$ into another Bloch state, $\ket{n \mathbf{k}}$, provided that the $n$-th band is nondegenerate along the reciprocal-space path prescribed by the acceleration theorem: $\mathbf{k}(t) = \mathbf{k}_0 + e \hbar^{-1} \mathbf{A}(t)$.
This adiabatic intraband motion is guaranteed in length-gauge and Houston-basis calculations, making these approaches a good choice when only ad hoc transition matrix elements are available~\cite{Apalkov_PRB_2012,Schiffrin_Nature_2013}.
In contrast, the correct adiabatic behavior is not explicitly encoded in the velocity-gauge equations of motion---it is implicit in the parameters of the model, demanding accurate energies and matrix elements.

The only matrix elements required in the velocity-gauge simulations are those of the momentum operator:
\begin{equation}
\label{eq:momentum_matrix_elements}
\mathbf{p}_{n m}(\mathbf{k}) = \frac{1}{\Omega} \int_{\mathrm{cell}} \de^3(\mathbf{r})\,
\psi_{n \mathbf{k}}^*(\mathbf{r}) \hat{\mathbf{p}} \psi_{m \mathbf{k}}(\mathbf{r}),
\end{equation}
where the integration is performed over a unit cell, $\Omega$ is the cell volume, and $\psi_{n \mathbf{k}}(\mathbf{r}) = \braket{\mathbf{r}}{n \mathbf{k}}$ are Bloch functions in the coordinate representation.
Indeed, let us search for time-dependent wave functions using the following ansatz:
\begin{equation}
\label{eq:WF_ansatz}
\ket{\psi_{\mathbf{k}}(t)} = \sum_m \alpha_{m \mathbf{k}}(t) \ket{m \mathbf{k}}.
\end{equation}
Inserting this ansatz into the time-dependent Schrödinger equation (TDSE), $\iu \hbar \partial_t \ket{\psi} = \hat{H}(t) \ket{\psi}$, and assuming that $\braket{m \mathbf{k'}}{n \mathbf{k}} = \delta_{m n} \delta(\mathbf{k}-\mathbf{k'})$, we obtain a system of coupled differential equations for the probability amplitudes:
\begin{equation}
\label{eq:TDSE2}
\iu \hbar \pdiff{\alpha_{q \mathbf{k}}}{t} = \epsilon_q(\mathbf{k}) \alpha_{q \mathbf{k}}
+ \frac{e}{m_0} \mathbf{A}(t) \cdot \sum_m \mathbf{p}_{q m}(\mathbf{k}) \alpha_{m \mathbf{k}}.
\end{equation}
In these equations, the medium properties are fully defined by $\epsilon_q(\mathbf{k})$ and $\mathbf{p}_{q m}(\mathbf{k})$.
Once the probability amplitudes are evaluated, the electric current density can be obtained according to
\begin{equation}
\label{eq:j}
\mathbf{J}(t) = \sum_{n \in \mathrm{occ}}
\int_{\mathrm{BZ}} \frac{\de^3\mathbf{k}}{(2 \pi)^3}\, f_{n}(\mathbf{k}) \mathbf{j}_{n \mathbf{k}}(t).
\end{equation}
Here, $f_{n}(\mathbf{k})$ is the Fermi factor of band $n$ in the initial (stationary) state.
The summation in Eq.~\eqref{eq:j} is carried out over all the bands that contain electrons in the unperturbed state (these would be valence bands for a dielectric), the integration is performed over the first Brillouin zone (BZ), and the contribution from crystal momentum $\mathbf{k}$ in initial band $n$ is given by
\begin{multline}
\label{eq:j_nk}
\mathbf{j}_{n \mathbf{k}}(t) = -\frac{e}{m_0} \Biggl( e \mathbf{A}(t) +\\
\re \biggl[ \sum_{i j} \left( \alpha_{i \mathbf{k}}^{(n)}(t) \right)^*
	\alpha_{j \mathbf{k}}^{(n)}(t) \mathbf{p}_{i j}(\mathbf{k})
 \biggr] \Biggr).
\end{multline}
In this equation, we have added a superscript to the probability amplitudes in order to denote the initial band:
$\alpha_{i \mathbf{k}}^{(n)}(t_\mathrm{min}) = \delta_{i n}$.
That is, in each initial state, only one band is occupied; by adding the contributions from all the initial states, as prescribed by Eq.~(5), we model a solid with occupied valence bands.
Once $\mathbf{J}(t)$ is evaluated, the polarization response, which is the key quantity in linear and nonlinear optics, is readily given by~\cite{King-Smith_PRB_1993,Resta_PRL_1998}
\begin{equation}
\mathbf{P}(t) = \int_{-\infty}^{t} \de t'\, \mathbf{J}(t').
\end{equation}


\section{Velocity-gauge corrections}
\label{sec:main_results}
In the previous section, we argued that velocity-gauge simulations tend to violate adiabaticity where adiabaticity is expected, which is why they demand high-quality band energies and matrix elements.
In this section, we derive analytical corrections that alleviate this issue in the case where a finite number of energy bands is the main source of discretization errors.
In the next section, we will demonstrate that our corrections work well even if the optical response is far from being adiabatic (that is, the polarization $\mathbf{P}$(t) is not a function of the electric field at time $t$).
In spite of that, we derive the corrections by considering an adiabatic limit.
It is worth clarifying this point before we start the derivation.
If velocity-gauge simulations require $N_{\mathrm{VG}}$ bands to produce physically meaningful results, while length-gauge simulations require only $N_{\mathrm{LG}} < N_{\mathrm{VG}}$ bands, then the bands that are unnecessary in the length gauge serve a rather numerical than physical purpose.
They are required in the velocity gauge to ensure that the basis set is reasonably complete.
After the interaction with an external field, the occupation of these highly excited states is negligibly small.
Even if the polarization response due to ``physically relevant'' bands is very nontrivial, the contributions due to very high conduction bands (and very low valence ones) are expected to be, in some sense, ``simple''.
We argue that this simplicity consists in the adiabaticity with respect to the vector potential: the difference between the exact current density, $\mathbf{J}(t)$, and that evaluated with a finite number of bands, $\mathbf{J}_N(t)$ is, approximately, a function of the vector potential, $\mathbf{A}$, at time $t$.
In this case, one may use any function $\mathbf{A}(t)$ that is particularly well suited for analytical calculations.
We chose
\begin{equation}
\mathbf{A}(t) = \re\left[  \mathbf{a} \ee^{\gamma t - \iu \omega_0 t}\right],
\end{equation}
where $\mathbf{a}$ is a constant vector (it is complex-valued unless $\mathbf{A}$ is linearly polarized), $\omega_0$ is a carrier-wave frequency, and $\gamma>0$ is a small parameter that controls how slowly the field is turned on.
The next step will be the evaluation of the current density using time-dependent perturbation theory, where we only consider a time interval where the external field may be viewed as a perturbation.
Once this step is accomplished, we will take the adiabatic limit: $\omega_0 \to 0$.
Here comes our key idea.
When the amplitude of the vector potential, $|\mathbf{a}|$, is fixed and its frequency is decreased, then the electric field decreases as well; consequently, we expect $\lim_{\gamma \to 0+} \lim_{\omega_0 \to 0} \mathbf{J}(t) \equiv 0$ for a dielectric and, more generally,
\begin{equation}\label{eq:key_limit}
\lim_{\gamma \to 0+} \lim_{\omega_0 \to 0} \mathbf{J}(t) \equiv \mathbf{J}_{\mathrm{ad}}\bigl(\mathbf{A}(t)\bigr)
\end{equation}
if there are partially occupied states in the unperturbed solid.
The adiabatic intraband current density, $\mathbf{J}_{\mathrm{ad}}$, is produced by charge carriers that change their crystal momentum according to the acceleration theorem, without leaving their initial bands:
\begin{equation}\label{eq:adiabatic_current}
\mathbf{J}_{\mathrm{ad}}(\mathbf{A}) = -\frac{e}{m_0} \sum_{n \in \mathrm{occ}}
\int_{\mathrm{BZ}} \frac{\de^3\mathbf{k}}{(2 \pi)^3}\, f_{n}(\mathbf{k})
\mathbf{p}_{n n}\left(\mathbf{k} + \frac{e}{\hbar} \mathbf{A}\right).
\end{equation}
Here, it is sufficient to carry out the sum over partially occupied bands.
Indeed, the contribution from empty bands is obviously zero, while for a fully occupied band, where $f_{n}(\mathbf{k}) \equiv 1$, the integral can be shown to be zero: $\int_{\mathrm{BZ}} \de^3\mathbf{k}\,\mathbf{p}_{n n}(\mathbf{k} + e \hbar^{-1} \mathbf{A}) = m_0 \hbar^{-1} \int_{\mathrm{BZ}} \de^3\mathbf{k}\,\nabla_{\mathbf{k}} \epsilon_n(\mathbf{k} + e \hbar^{-1} \mathbf{A})  = 0$.

In the velocity gauge, the equations of motion do not automatically satisfy Eq.~\eqref{eq:key_limit}.
We will soon see that this equation is, in general, violated by truncating the basis set, which is an unavoidable operation for numerical calculations.
However, knowing that Eq.~\eqref{eq:key_limit} holds in the ideal case, we arrive at a simple recipe for adiabatic corrections to the electric current density evaluated with a finite number of bands:
\begin{equation}\label{eq:recipe}
\Delta\mathbf{J}_N = \lim_{\gamma \to 0+} \lim_{\omega_0 \to 0}
\bigl[ \mathbf{J}(t) - \mathbf{J}_N(t) \bigr] =
\mathbf{J}_{\mathrm{ad}} - \lim_{\gamma \to 0+} \lim_{\omega_0 \to 0} \mathbf{J}_N.
\end{equation}
In the adiabatic limit, $\Delta\mathbf{J}_N$ is a function of $\mathbf{A}$.
In the rest of this section, we will be searching for a decomposition of $\Delta\mathbf{J}_N$ into the powers of the vector potential.
As a first step, we decompose the intraband adiabatic current:
\begin{equation}
	\mathbf{J}_{\mathrm{ad}}(\mathbf{A}) = \mathbf{J}_{\mathrm{ad}}^{(0)} + \mathbf{J}_{\mathrm{ad}}^{(1)}(\mathbf{A}) + \mathbf{J}_{\mathrm{ad}}^{(2)}(\mathbf{A}) + \mathbf{J}_{\mathrm{ad}}^{(3)}(\mathbf{A}) + \ldots
\end{equation} 
The first two terms in this decomposition are
\begin{align}
\label{eq:J_ad0}
	\mathbf{J}_{\mathrm{ad}}^{(0)} &= -\frac{e}{m_0} \sum_{n \in \mathrm{occ}}
	\int_{\mathrm{BZ}} \frac{\de^3\mathbf{k}}{(2 \pi)^3}\, f_{n}(\mathbf{k})
	\mathbf{p}_{n n}\left(\mathbf{k}\right),\\
\label{eq:J_ad1}
   \mathbf{J}_{\mathrm{ad}}^{(1)} &= -\frac{e^2}{\hbar m_0} \sum_{n \in \mathrm{occ}}
	\int_{\mathrm{BZ}} \frac{\de^3\mathbf{k}}{(2 \pi)^3}\, f_{n}(\mathbf{k})
	\scalarproduct{\mathbf{A}}{\nabla_{\mathbf{k}}} \mathbf{p}_{n n}(\mathbf{k}).
\end{align}
The other terms are straightforward to obtain by continuing the Taylor expansion.

We proceed with a detailed derivation of the first-order correction to the current density.
In first-order perturbation theory, the solution of Eq.~\eqref{eq:TDSE2} with an initial condition
$\lim_{t \to -\infty} \alpha_{q \mathbf{k}}^{(n)}(t) \exp[\frac{\iu}{\hbar} \epsilon_q(\mathbf{k}) t] = \delta_{q n}$
is
\begin{multline}\label{eq:first-order_alpha}
\alpha_{q \mathbf{k}}^{(n)}(t) \approx
\exp\left[-\frac{\iu}{\hbar} \epsilon_n(\mathbf{k}) t\right] \delta_{q n} -\\
\frac{e}{2 \hbar m_0}
\exp\left[\left(\gamma - \frac{\iu}{\hbar} \epsilon_n(\mathbf{k})\right) t\right]
\times\\
 \Biggl(
\frac{\ee^{- \iu \omega_0 t} \scalarproduct{\mathbf{p}_{q n}(\mathbf{k})}{\mathbf{a}}}
  {\omega_{q n}(\mathbf{k}) - \omega_0 - \iu \gamma} +
\frac{\ee^{\iu \omega_0 t}
	\scalarproduct{\mathbf{p}_{q n}(\mathbf{k})}{\mathbf{a}^*}}
{\omega_{q n}(\mathbf{k}) + \omega_0 - \iu \gamma}
\Biggr),
\end{multline}
where we have introduced transition frequencies:
\begin{equation}\label{eq:transition_frequencies}
  \omega_{q n}(\mathbf{k}) = \left(\epsilon_q(\mathbf{k}) - \epsilon_n(\mathbf{k}) \right) / \hbar.
\end{equation}
From now on, we will omit, for brevity, $\mathbf{k}$ in the arguments of the transition frequencies and momentum matrix elements.
Substituting Eq.~\eqref{eq:first-order_alpha} into Eq.~\eqref{eq:j_nk}, neglecting the terms quadratic with respect to $\mathbf{a}$, reintroducing $\mathbf{A}(t)$ by substituting $\mathbf{a}^*$ with $2 \mathbf{A}(t) \ee^{-\gamma t - \iu \omega_0 t} - \mathbf{a} \ee^{-2 \iu \omega_0 t}$, and taking the limit $\omega_0 \to 0$, we obtain
\begin{multline*}
\lim_{\omega_0 \to 0} \mathbf{j}_{n \mathbf{k}} =
-\frac{e}{m_0} \mathbf{p}_{n n} +
\frac{e^2}{m_0} \mathbf{A} -
\frac{2 e^2}{\hbar m_0^2} \times\\
\sum_i 
\frac{
	\re \left[
	  ( \omega_{i n} + \iu \gamma) \scalarproduct{\mathbf{p}_{i n}}{\mathbf{A}} \mathbf{p}_{n i}
	\right]}{\gamma^2 + \omega_{i n}^2}.
\end{multline*}
Before we take the limit $\gamma \to 0+$, we notice that those terms under the sum where $\omega_{i n} = 0$ vanish in this limit.
In the rest of this paper, the notation $\sum_{i \ne n}$ will imply that not only the term $i=n$ but also all other terms where $\omega_{i n}=0$, if present, must be omitted (otherwise, a zero would appear in the denominator).
With this in mind, we take the limit $\gamma \to 0+$, use Eq.~\eqref{eq:j}, and write the sum of the zero- and first-order approximations to the adiabatic current density as
\begin{multline}\label{eq:first-order_current}
\lim_{\gamma \to 0+} \lim_{\omega_0 \to 0} \mathbf{J} \approx
- \frac{e}{m_0} \sum_{n \in \mathrm{occ}}
\int_{\mathrm{BZ}} \frac{\de^3\mathbf{k}}{(2 \pi)^3}\, f_{n}(\mathbf{k}) \mathbf{p}_{n n} -\\
\frac{e^2}{m_0} \sum_{n \in \mathrm{occ}}
\int_{\mathrm{BZ}} \frac{\de^3\mathbf{k}}{(2 \pi)^3}\, f_{n}(\mathbf{k}) \Biggl\{
\mathbf{A} -
\frac{2}{m_0}
\sum_{i \ne n}\frac{\re\Bigl[\scalarproduct{\mathbf{p}_{n i}}{\mathbf{A}} \mathbf{p}_{i n}\Bigr]}
{\hbar \omega_{i n}}
\Biggr\}.
\end{multline}
The first term on the right-hand side of this expression, representing the electric current that flows without external field, is identical with the zero-order expansion term of the intraband adiabatic current, Eq.~\eqref{eq:J_ad0}.
The two terms cancel when we evaluate $\Delta\mathbf{J}_N$ according to Eq.~\eqref{eq:recipe}.
In the rest of Eq.~\eqref{eq:first-order_current}, we recognize the Thomas-Reiche-Kuhn sum rule in the form applicable to solids~\cite{Luttinger_PR_1955}.
For our purposes, it is convenient to write this rule as
\begin{equation}\label{eq:TRK_sum_rule}
\frac{2}{m_0}
\sum_{i \ne n}\frac{\bigl|\scalarproduct{\mathbf{e}_\beta}{\mathbf{p}_{i n}}\bigr|^2}{\hbar \omega_{i n}}
= 1 - \hbar \pdiff{\scalarproduct{\mathbf{e}_\beta}{\mathbf{p}_{n n}}}{k_{\beta}},
\end{equation}
where $\beta \in \{x,y,z\}$ is a Cartesian coordinate, and $\mathbf{e}_\beta$ is a unit vector along the corresponding coordinate axis.
Remembering Eq.~\eqref{eq:J_ad1}, we see that if the summation in Eq.~\eqref{eq:first-order_current} is performed over the entire infinite manifold of bands, then the requirement encoded in Eq.~\eqref{eq:key_limit} leads to the Thomas-Reiche-Kuhn sum rule.
If we truncate the basis set, we violate the sum rule.

Following Eq.~\eqref{eq:recipe}, we obtain the following first-order correction to the electric current density obtained by solving the TDSE numerically with a finite number of bands:
\begin{multline}\label{eq:first-order_correction}
\Delta\mathbf{J}^{(1)} = \mathbf{J}_{\mathrm{ad}}^{(1)}(\mathbf{A}) +
\frac{e^2}{m_0} \sum_{n \in \mathrm{occ}}
\int_{\mathrm{BZ}} \frac{\de^3\mathbf{k}}{(2 \pi)^3}\, f_n(\mathbf{k}) \Biggl\{
\mathbf{A} -\\
\frac{2}{m_0}
\sum_{i \ne n}^{N}\frac{\re\Bigl[\scalarproduct{\mathbf{p}_{n i}}{\mathbf{A}} \mathbf{p}_{i n}\Bigr]}{\hbar \omega_{i n}}
\Biggr\} \\=
\frac{e^2}{m_0} \sum_{n \in \mathrm{occ}}
\int_{\mathrm{BZ}} \frac{\de^3\mathbf{k}}{(2 \pi)^3}\, f_n(\mathbf{k}) \Biggl\{
\mathbf{A} -
\frac{1}{\hbar}
\scalarproduct{\mathbf{A}}{\nabla_{\mathbf{k}}} \mathbf{p}_{n n} -\\
\frac{2}{m_0}
\sum_{i \ne n}^{N}\frac{\re\Bigl[\scalarproduct{\mathbf{p}_{n i}}{\mathbf{A}} \mathbf{p}_{i n}\Bigr]}{\hbar \omega_{i n}}
\Biggr\}.
\end{multline}
Here and in the following, summation is performed only over those bands that  $\mathbf{J}_N(t)$ was evaluated with.
As a reminder that the basis is truncated, we have replaced $\sum_{i \ne n}$ with $\sum_{i \ne n}^{N}$.

For a dielectric, where $\mathbf{J}_{\mathrm{ad}}(\mathbf{A}) \equiv 0$, 
$f_n(\mathbf{k}) = 1$ for valence bands, and $f_n(\mathbf{k}) = 0$ for conduction bands,
we can write Eq.~\eqref{eq:first-order_correction} as
\begin{multline}\label{eq:first-order_correction_for_dielectrics}
\Delta\mathbf{J}_{\mathrm{dielectric}}^{(1)} =
\frac{e^2}{m_0}
\int_{\mathrm{BZ}} \frac{\de^3\mathbf{k}}{(2 \pi)^3}\, \Biggl\{
N_{\mathrm{VB}} \mathbf{A} -\\
\frac{2}{m_0} \sum_{n \in \mathrm{VB}}
\sum_{i \ne n}^{N}\frac{\re\Bigl[\scalarproduct{\mathbf{p}_{n i}}{\mathbf{A}} \mathbf{p}_{i n}\Bigr]}{\hbar \omega_{i n}}
\Biggr\}.
\end{multline}
Here, $N_{\mathrm{VB}} = \sum_{n \in \mathrm{occ}} f_n$ is the number of valence bands in a simulation to which the correction is applied; we denote this set of bands as ``VB''.
To emphasize that we consider the case where all the valence bands are initially occupied, while the conduction bands are empty, we have replaced $n \in \mathrm{occ}$ with $n \in \mathrm{VB}$.
The summation over $i$ goes over all the valence and conduction bands, except band $n$ and those terms where $\omega_{i n}(\mathbf{k})=0$; however, it is easy to see that the terms with $i \in \mathrm{VB}$ cancel. 

For dielectrics with a crystal symmetry that demands that the electric current must flow along the laser polarization (e.g. cubic symmetry), Eq.~\eqref{eq:first-order_correction_for_dielectrics} is equivalent to substituting the actual number of valence bands, $N_{\mathrm{VB}}$, with an effective one, $N_{\mathrm{VB}}^{\mathrm{eff}}$.
Indeed, by multiplying both sides of Eq.~\eqref{eq:first-order_correction_for_dielectrics} with $\mathbf{A}$ and relying on $\Delta\mathbf{J}^{(1)} \parallel \mathbf{A}$, we obtain
\begin{multline*}
\Delta \mathbf{J}_{\mathrm{dielectric}}^{(1)} =
\mathbf{A}(t) \frac{e^2}{m_0} \int_{\mathrm{BZ}} \frac{\de^3\mathbf{k}}{(2 \pi)^3}\,
\Biggl\{
N_{\mathrm{VB}} -\\
\frac{2}{m_0}
\sum_{n \in \mathrm{occ}}
\sum_{i \ne n}^{N}\frac{\Bigl|\mathbf{e}_\mathrm{L} \cdot \mathbf{p}_{i n} \Bigr|^2}{\hbar \omega_{i n}}
\Biggr\},
\end{multline*}
where $\mathbf{e}_\mathrm{L}$ is a unit vector pointing along $\mathbf{A}$.
Comparing this expression with Eq.~\eqref{eq:j_nk}, we see that our correction gives the same effect as substituting, in Eq.~\eqref{eq:j_nk}, $\mathbf{A}(t)$ with $\mathbf{A}(t) N_{\mathrm{VB}}^{\mathrm{eff}} / N_{\mathrm{VB}}$, where
\begin{equation}\label{NB_eff}
N_{\mathrm{VB}}^{\mathrm{eff}} = 
\frac{2 \Omega}{m_0}
\sum_{n \in \mathrm{VB}} \sum_{i \ne n}^{N}
\int_{\mathrm{BZ}} \frac{\de^3\mathbf{k}}{(2 \pi)^3}\,
\frac{\Bigl|\mathbf{e}_\mathrm{L} \cdot \mathbf{p}_{i n} \Bigr|^2}{\hbar \omega_{i n}}.
\end{equation}
The concept of an effective number of valence electrons, $N_{\mathrm{VB}}^{\mathrm{eff}}$, is well known~\cite{Philipp_PR_1963,Smith_PRB_1978,Wooten_2013_N_eff}, but it is not as widely known that $N_{\mathrm{VB}}^{\mathrm{eff}}$ can be used to adjust the polarization response obtained with an insufficient number of bands.
Also, in the next section, we will demonstrate that the applicability of this concept is not limited to the linear response---Eq.~\eqref{eq:first-order_correction} works surprisingly well even if the polarization response is very nonlinear.

Note that $N_{\mathrm{VB}}$ does not depend on the field magnitude; consequently, our corrections retain their importance in the weak-field limit.
In other words, no matter how weak the electric field is, one has to use a sufficient number of bands to either satisfy the Thomas-Reiche-Kuhn sum rule or be able to compensate for its violation. 

The derivation of higher-order corrections is similar to the one presented above, but it is much more laborious.
Therefore, we only present the final results.
For the second-order correction, we get
\begin{multline}\label{eq:second-order_correction}
  \Delta\mathbf{J}^{(2)} =  \Delta\mathbf{J}_{\mathrm{ad}}^{(2)} +\\
  \frac{e^3}{\hbar^2 m_0^3}
  \sum_{n \in \mathrm{occ}} \int_{\mathrm{BZ}} \frac{\de^3\mathbf{k}}{(2 \pi)^3}\,
  f_n(\mathbf{k})
  \Biggl\{ \Biggl(
  \sum_{i \ne n}^{N}\sum_{j \ne n}^{N} \frac{1}{\omega_{i n} \omega_{j n}} \\
  \times \Biggl[
  \mathbf{p}_{j n} \scalarproduct{\mathbf{p}_{i j}}{\mathbf{A}}
  \scalarproduct{\mathbf{p}_{n i}}{\mathbf{A}} +\\ 
  \scalarproduct{\mathbf{p}_{j n}}{\mathbf{A}}
  \biggl(\mathbf{p}_{i j} \scalarproduct{\mathbf{p}_{n i}}{\mathbf{A}}+
  \mathbf{p}_{n i} \scalarproduct{\mathbf{p}_{i j}}{\mathbf{A}}
  \biggr) \Biggr] \Biggr)
  -\\
  \sum_{i \ne n}^{N}\frac{2 \scalarproduct{\mathbf{p}_{n n}}{\mathbf{A}} \re\left[
  	\mathbf{p}_{n i} \scalarproduct{\mathbf{p}_{i n}}{\mathbf{A}} \right] +
  	\mathbf{p}_{n n} \left|\mathbf{p}_{n i} \cdot \mathbf{A}\right|^2}{\omega_{i n}^2}
  \Biggr\}.
\end{multline}
This expression necessarily equals zero if $\hat{H}^{(0)}$ is symmetric with respect to time reversal and $f_{n}(-\mathbf{k}) = f_{n}(\mathbf{k})$.

Finally, for the third-order correction, we obtain
\begin{widetext}
\begin{multline}\label{eq:third-order_correction}
  \Delta\mathbf{J}^{(3)} = \Delta\mathbf{J}_{\mathrm{ad}}^{(3)} + \frac{e^4}{\hbar^3 m_0^4}
  \sum_{n \in \mathrm{occ}} \int_{\mathrm{BZ}} \frac{\de^3\mathbf{k}}{(2 \pi)^3}\,
  f_n(\mathbf{k})
  \Biggl\{
  -\sum_{i \ne n}^{N}\sum_{j \ne n}^{N}\sum_{k \ne n}^{N} \Biggl(
  \frac{1}{\omega_{i n} \omega_{j n} \omega_{k n}}\re \Biggl[
  \scalarproduct{\mathbf{p}_{j i}}{\mathbf{A}} \times\\
  \biggl(\scalarproduct{\mathbf{p}_{n k}}{\mathbf{A}}
  \Bigl[\mathbf{p}_{i n} \scalarproduct{\mathbf{p}_{k j}}{\mathbf{A}} +
  \mathbf{p}_{k j} \scalarproduct{\mathbf{p}_{i n}}{\mathbf{A}}\Bigr] +
  \mathbf{p}_{n k} \scalarproduct{\mathbf{p}_{i n}}{\mathbf{A}} 
  \scalarproduct{\mathbf{p}_{k j}}{\mathbf{A}} \biggr) +\\
  \mathbf{p}_{j i} \scalarproduct{\mathbf{p}_{i n}}{\mathbf{A}} 
  \scalarproduct{\mathbf{p}_{k j}}{\mathbf{A}} \scalarproduct{\mathbf{p}_{n k}}{\mathbf{A}}
  \Biggr] \Biggr) +
  \sum_{i \ne n}^{N}\sum_{j \ne n}^{N} \Biggl(
  \frac{\omega_{i n}+\omega_{j n}}{2 \omega_{i n}^2 \omega_{j n}^2}\Biggl[
  \mathbf{p}_{j i} \scalarproduct{\mathbf{p}_{n n}}{\mathbf{A}} \scalarproduct{\mathbf{p}_{i n}}{\mathbf{A}} \scalarproduct{\mathbf{p}_{n j}}{\mathbf{A}} +\\
  \mathbf{p}_{j n} \scalarproduct{\mathbf{p}_{n i}}{\mathbf{A}}
  \Bigl[\scalarproduct{\mathbf{p}_{i n}}{\mathbf{A}} \scalarproduct{\mathbf{p}_{n j}}{\mathbf{A}}+\scalarproduct{\mathbf{p}_{n n}}{\mathbf{A}} \scalarproduct{\mathbf{p}_{i j}}{\mathbf{A}}\Bigr] +\\
  \scalarproduct{\mathbf{p}_{j i}}{\mathbf{A}} \Bigl[
  \scalarproduct{\mathbf{p}_{n n}}{\mathbf{A}} \Bigl(\mathbf{p}_{i n} \scalarproduct{\mathbf{p}_{n j}}{\mathbf{A}} + \mathbf{p}_{n j} \scalarproduct{\mathbf{p}_{i n}}{\mathbf{A}}\Bigr) +
  \mathbf{p}_{n n} \scalarproduct{\mathbf{p}_{i n}}{\mathbf{A}} \scalarproduct{\mathbf{p}_{n j}}{\mathbf{A}} \Bigr] +\\ 
  \scalarproduct{\mathbf{p}_{j n}}{\mathbf{A}} \Biggl(
  \mathbf{p}_{n i} \Bigl(\scalarproduct{\mathbf{p}_{i n}}{\mathbf{A}} \scalarproduct{\mathbf{p}_{n j}}{\mathbf{A}} +
  \scalarproduct{\mathbf{p}_{n n}}{\mathbf{A}} \scalarproduct{\mathbf{p}_{i j}}{\mathbf{A}} \Bigr) +\\
  \scalarproduct{\mathbf{p}_{n i}}{\mathbf{A}} \Bigl[ \mathbf{p}_{i j} \scalarproduct{\mathbf{p}_{n n}}{\mathbf{A}} +
  \mathbf{p}_{i n} \scalarproduct{\mathbf{p}_{n j}}{\mathbf{A}}+\mathbf{p}_{n j} \scalarproduct{\mathbf{p}_{i n}}{\mathbf{A}}+\mathbf{p}_{n n} \scalarproduct{\mathbf{p}_{i j}}{\mathbf{A}}
  \Bigr] \Biggr) \Biggr] \Biggr) -\\
  \scalarproduct{\mathbf{p}_{n n}}{\mathbf{A}}
  \sum_{i \ne n}^{N}\frac{\mathbf{p}_{n i} \scalarproduct{\mathbf{p}_{n n}}{\mathbf{A}} \scalarproduct{\mathbf{p}_{i n}}{\mathbf{A}} +
  	\scalarproduct{\mathbf{p}_{n i}}{\mathbf{A}}
  	\Bigl(\mathbf{p}_{i n} \scalarproduct{\mathbf{p}_{n n}}{\mathbf{A}}+2 \mathbf{p}_{n n} \scalarproduct{\mathbf{p}_{i n}}{\mathbf{A}}\Bigr)}{\omega_{i n}^3} 
  \Biggr\}.
\end{multline}
\end{widetext}
Even though it is a rather lengthy expression, we will demonstrate that it may be very useful.

\section{A one-dimensional demonstration of the method}
\label{sec:example}
We derived our adiabatic corrections in three spatial dimensions; nevertheless, we demonstrate their power by means of one-dimensional simulations.
On the one hand, it is much easier to control and achieve numerical convergence in one-dimensional calculations; on the other hand, the equations derived in the previous section become more manageable once we replace all vector quantities with scalar ones.
Also, we further simplify the equations by considering a dielectric and eliminating $f_n(\mathbf{k})$.
We write our corrections to the current density as an expansion in powers of the vector potential:
\begin{equation}\label{eq:1D_Delta_j}
\Delta J(t) = \sum_q c_q A^q(t).
\end{equation}
In Eqs.~\eqref{eq:j}, \eqref{eq:first-order_correction}, \eqref{eq:second-order_correction}, and \eqref{eq:third-order_correction}, we replace $\de^3 \mathbf{k}/(2 \pi)^3$ with $\de k/(2 \pi)$, substitute scalar quantities for the vector ones, and, after some simplifications, obtain
\begin{equation}\label{eq:1D_1st-order_correction}
  c_1 =
  \frac{e^2}{m_0} \int_{\mathrm{BZ}} \frac{\de k}{2 \pi}\,
  \Biggl\{
  N_{\mathrm{VB}} -
  \frac{2}{\hbar m_0}
  \sum_{n \in \mathrm{VB}}
  \sum_{i \ne n}^{N} \frac{|p_{i n}|^2}{\omega_{i n}}
  \Biggr\},
\end{equation}
\begin{multline}\label{eq:1D_2nd-order_correction}
c_2 = \frac{3 e^3}{\hbar^2 m_0^3}
\sum_{n \in \mathrm{VB}} \int_{\mathrm{BZ}} \frac{\de k}{2 \pi}\,
\Biggl\{ \sum_{i \ne n}^{N}\sum_{j \ne n}^{N} \frac{p_{i j} p_{n i}
p_{j n}}{\omega_{i n} \omega_{j n}} -\\
 p_{n n} \sum_{i \ne n}^{N} \frac{\left|
p_{i n}\right| ^2}{\omega_{i n}^2}\Biggr\},
\end{multline}
and
\begin{multline}\label{eq:1D_3rd-order_correction}
c_3 = -\frac{4 e^4}{\hbar^3 m_0^4} \times\\
\sum_{n \in \mathrm{VB}} \int_{\mathrm{BZ}} \frac{\de k}{2 \pi}\,
\Biggl\{
\sum_{i \ne n}^{N}\sum_{j \ne n}^{N}\sum_{\ell \ne n}^{N} \frac{\re[p_{j i} p_{i n}
		p_{\ell j} p_{n \ell}]}{\omega_{i n} \omega_{j n}
		\omega_{\ell n}} -\\
\sum_{i \ne n}^{N}\sum_{j \ne n}^{N} \frac{(\omega_{i n}+\omega_{j n})
\left(\frac{\left| p_{i n} p_{j n}\right|^2}{2} + p_{n n}
\re[p_{i j} p_{n i} p_{j n}]\right)}{\omega_{i n}^2 \omega_{j n}^2} +\\
p_{n n}^2 \sum_{i \ne n}^{N} \frac{\left| p_{i n}\right|^2}{\omega_{i n}^3}
\Biggr\}.
\end{multline}
These coefficients depend only on the stationary electronic structure of the solid, and they do not depend on the laser pulse.
Also, evaluating the coefficients, we have dropped the adiabatic intraband current density, which is supposed to be zero for dielectrics but may acquire nonzero values when the integral in Eq.~\eqref{eq:adiabatic_current} is evaluated numerically.
We found that these values were so small that accounting for them had a negligible effect on the results presented in this section.

In the following, we use atomic units ($\hbar = e = m_0 = 1$) unless specified otherwise.
For our numerical demonstration, we employ a stationary Hamiltonian, $\hat{H}^{(0)} = -\frac{1}{2} \frac{\de^2}{\de x^2} + V(x)$, with a noncentrosymmetric lattice potential:
\begin{multline}\label{eq:lattice_potential}
V(x) = \sum_q \Biggl\{ -2.2\,\text{sech}^2\bigl(0.9 (x - q a)\bigr) +\\
0.01 \sin\left(\frac{2 \pi (x - q a)}{a}\right)
\Biggr\}.
\end{multline}
Here, $a = 9.45\ \text{at.\,u.} = 5$~{\AA} is the lattice constant.
Evaluating the band structure, we Fourier-expanded $V(x)$ into 81 plane waves.
We regard the lowest two energy bands of this potential as the valence ones.
The energy gap between the second and the third bands is equal to $\epsilon_3(0) - \epsilon_2(0) = 9$~eV, which is close to the band gap of quartz.
$\hat{H}^{(0)}$ is symmetric with respect to time reversal; therefore, up to round-off errors, $c_2$ in Eq.~\eqref{eq:1D_2nd-order_correction} evaluates to zero.

We numerically solved the TDSE for 61 uniformly spaced crystal momenta using only those stationary states that had an energy below a certain level.
We measure this level from the bottom of the lowest conduction band and refer to it as the cut-off energy, $\epsilon_{\mathrm{cut}}$.
In other words, the subset of eigenstates of $\hat{H}^{(0)}$ that we use to solve the TDSE consists of all the Bloch states with energies $\epsilon_q(k) \le\epsilon_{\mathrm{cut}} + \epsilon_3(0)$.

We used the following expression for the vector potential of the external field:
\begin{equation}
\label{eq:laser_pulse}
A(t) = -\theta(\tau_\mathrm{L}-|t|)
\frac{E_0}{\omega_0}
\cos^4\left(\frac{\pi t}{2 \tau_\mathrm{L}}\right)
\sin(\omega_0 t).
\end{equation}
Here, $E_0$ is the amplitude of the electric field, $\omega_0$ is the central frequency of the laser pulse, $\theta(x)$ is the Heaviside step function, and $\tau_\mathrm{L}$ is related to the full width at half maximum (FWHM) of $A^2(t)$ by 
\begin{equation} \label{eq:FWHM}
 \mbox{FWHM} = \frac{4 \tau_\mathrm{L}}{\pi} \arccos\left(2^{-1/8}\right).
\end{equation}
We chose the central wavelength of the laser pulse to be 750~nm ($\omega_0 = 2.51\ \text{fs}^{-1} = 0.06$~at.\,u.; $\hbar\omega_0 = 1.65$~eV) and set the FWHM to 4~fs ($\tau_\mathrm{L}=7.66\ \text{fs} = 317$~at.\,u.).

\begin{figure}[t]
	\begin{tabular}{p{0.9\columnwidth} p{0pt}}
		\vspace{0pt} \includegraphics[width=0.9\columnwidth]{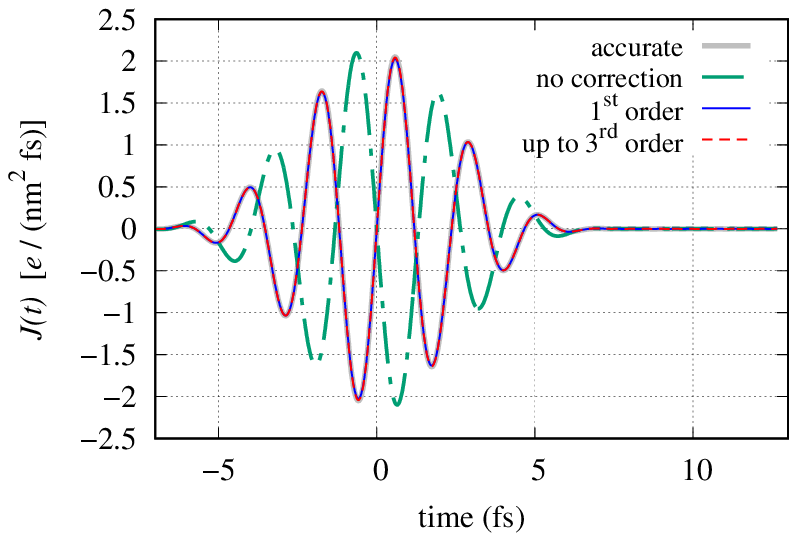} &
		\vspace{1mm} \hspace{-0.95\columnwidth}
		\textbf{(a)}
	\end{tabular}\\[-5mm]
	\begin{tabular}{p{0.9\columnwidth} p{0pt}}
		\vspace{0pt} \includegraphics[width=0.9\columnwidth]{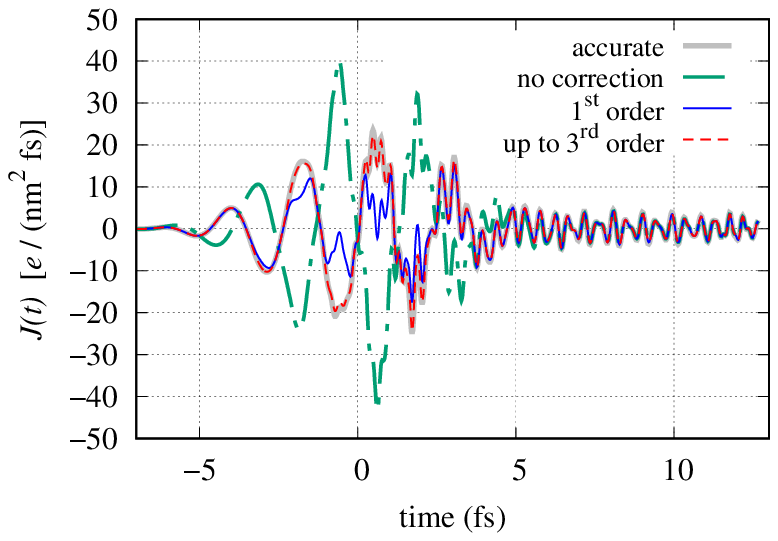} &
		\vspace{1mm} \hspace{-0.95\columnwidth}
		\textbf{(b)}
	\end{tabular}
	\caption{The electric current density evaluated for pulses with peak electric fields of (a) 0.1~V/{\AA} and (b) 1~V/{\AA}.
	The grey curves show current densities evaluated with a cut-off energy that was sufficient for convergence without corrections (2391.4~eV).
	The other curves represent outcomes of TDSE simulations with $\epsilon_{\mathrm{cut}}=25$~eV: without corrections (green dash-dotted curve), with the first-order correction (solid blue curve), and with the correction terms up to the third order (red dashed curve).}
	\label{fig:selected_currents}
\end{figure}

\begin{figure}[t]
	\begin{tabular}{p{0.9\columnwidth} p{0pt}}
		\vspace{0pt} \includegraphics[width=0.9\columnwidth]{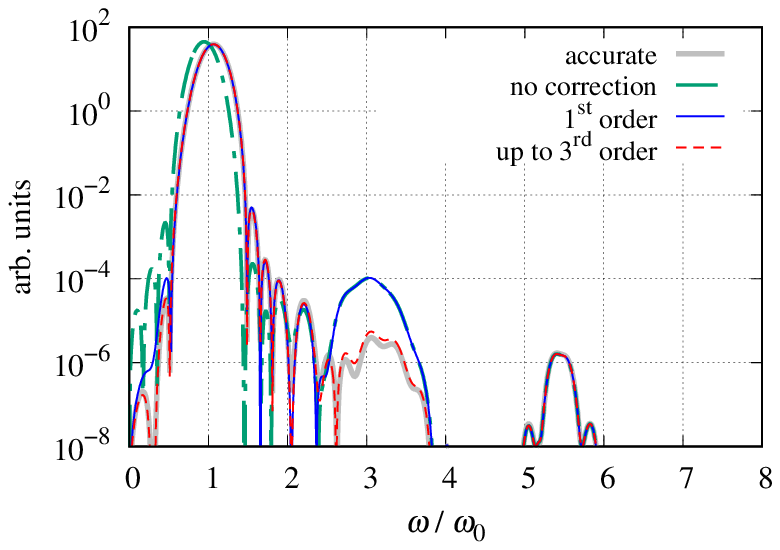} &
		\vspace{1mm} \hspace{-0.95\columnwidth}
		\textbf{(a)}
	\end{tabular}\\[-5mm]
	\begin{tabular}{p{0.9\columnwidth} p{0pt}}
		\vspace{0pt} \includegraphics[width=0.9\columnwidth]{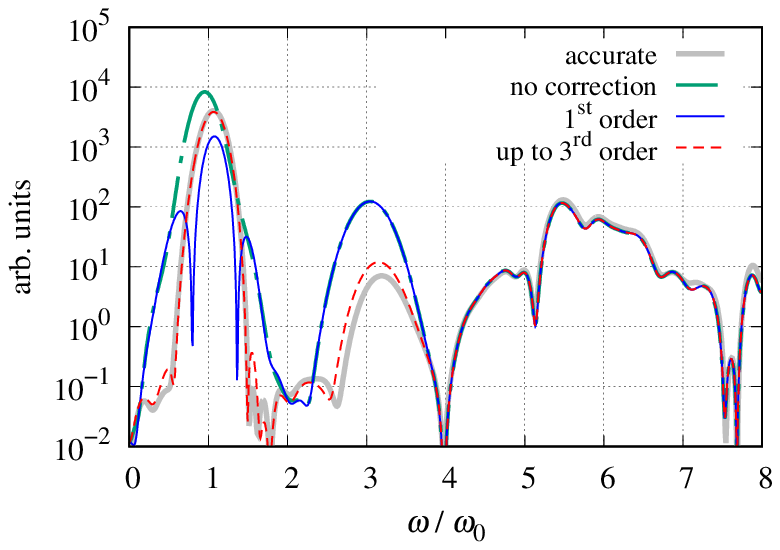} &
		\vspace{1mm} \hspace{-0.95\columnwidth}
		\textbf{(b)}
	\end{tabular}
	\caption{The spectra of the current densities shown if Fig.~\ref{fig:selected_currents}: (a) $E_0 = 0.1$~V/{\AA}, (b) $E_0 = 1$~V/{\AA}.}
	\label{fig:current_spectra}
\end{figure}

In Fig.~\ref{fig:selected_currents}, we compare electric current densities evaluated in different approximations for two peak values of the electric field: $E_0 = 0.1$~V/{\AA} for panel (a) and $E_0 = 1$~V/{\AA} for panel (b).
The spectra of these current densities are compared in Fig.~\ref{fig:current_spectra}.
We multiplied the current density with a constant factor (0.96333) that made the refractive index at the central laser wavelength equal to 1.47, which is the refractive index of thin-film fused silica at 750~nm \cite{Gao_JEOS_2013}.
This procedure allows us to label the $y$-axis of Fig.~\ref{fig:selected_currents} with units of $e/(\text{nm}^2\,\text{fs})$.
The current density labeled as ``accurate'' was obtained with $\epsilon_{\mathrm{cut}}=2391.4$~eV, which corresponds to using 40 energy bands.
We verified that, for this large value of $\epsilon_{\mathrm{cut}}$, corrections are unnecessary, so we did not apply them.
The other results in Fig.~\ref{fig:selected_currents} were evaluated for $\epsilon_{\mathrm{cut}}=25$~eV, which corresponds to using just the lowest three conduction bands of our one-dimensional model.
Without corrections, the velocity gauge fails dramatically for this low value of $\epsilon_{\mathrm{cut}}$.
However, already the first-order correction, specified by Eqs.~\eqref{eq:1D_Delta_j} and \eqref{eq:1D_1st-order_correction}, greatly improves the accuracy.
In Fig.~\ref{fig:selected_currents}(a), the curves representing the accurate and corrected current densities are virtually indistinguishable.
When we increase the peak electric field to $E_0 = 1$~V/{\AA}, we see significant deviations between the accurate $J(t)$ (thick gray curve) and that obtained in the simulation with $\epsilon_{\mathrm{cut}}=25$~eV, followed by applying the first-order correction (solid blue curve).
These deviations disappear as soon as we add the third-order correction to the first-order one (red dashed curve).

In the frequency domain, the first-order correction only affects the spectral range at the central laser frequency, while the third-order correction affects the spectral components at $\omega_0$  and $3 \omega_0$.
This is clearly seen in Fig.~\ref{fig:current_spectra}, where we also observe that high frequencies ($\omega \gtrsim 4 \omega_0$) are much less demanding to the number of bands than the low frequencies are.
Note that the current density in Figs.~\ref{fig:selected_currents}(b) and Figs.~\ref{fig:current_spectra}(b) shows signatures of highly nonlinear processes.
In particular, the fast oscillations in Figs.~\ref{fig:selected_currents}(b) emerge due to the multiphoton absorption~\cite{Korbman_NJP_2013} (in our case, the ratio of the band gap to the photon energy is equal to 5.4).
On average, $6.6 \times 10^{-4}$ electrons per unit cell are excited to conduction bands by the pulse with $E_0 = 1$~V/{\AA}; this value grows up to $1.3 \times 10^{-2}$ for $E_0 = 1.5$~V/{\AA}.
For $t > \tau_\mathrm{L}=7.66\ \text{fs}$, the three curves obtained with $\epsilon_{\mathrm{cut}}=25$~eV coincide because the vector potential is zero, and so are the correction terms.
The fact that these curves are very close to our most accurate current density means that a negligible number of electrons is excited to energies above $\epsilon_{\mathrm{cut}}$.
This is expected because bands that carry residual population are responsible for processes more complex than those that our adiabatic corrections account for.

\begin{figure}[!htb]
	\begin{tabular}{p{0.9\columnwidth} p{0pt}}
		\vspace{0pt} \includegraphics[width=0.9\columnwidth]{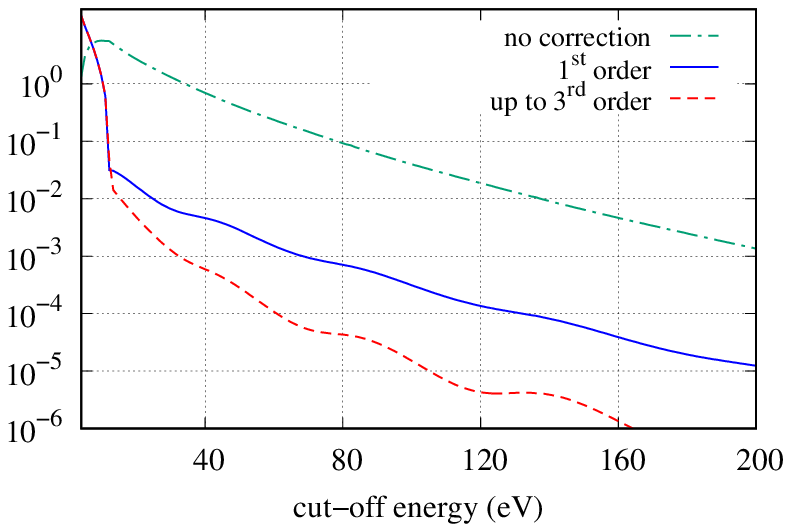} &
		\vspace{1mm} \hspace{-0.95\columnwidth}
		\textbf{(a)}
	\end{tabular}\\[-5mm]
	\begin{tabular}{p{0.9\columnwidth} p{0pt}}
		\vspace{0pt} \includegraphics[width=0.9\columnwidth]{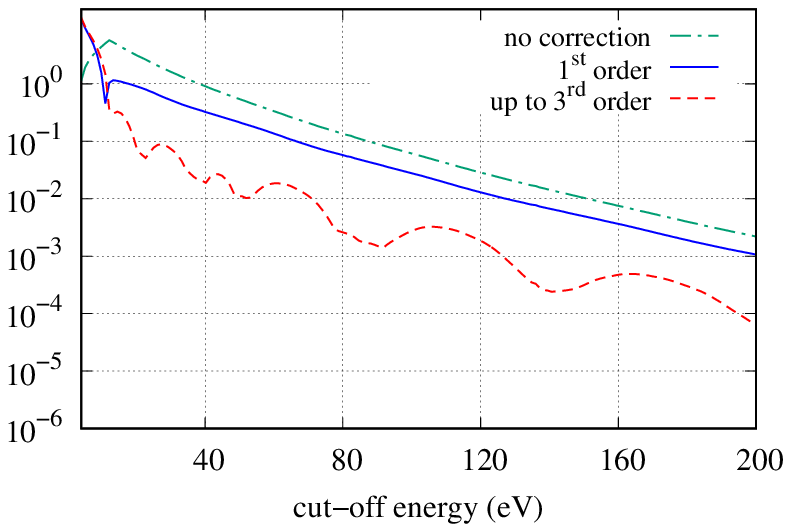} &
		\vspace{1mm} \hspace{-0.95\columnwidth}
		\textbf{(b)}
	\end{tabular}
	\caption{Discrepancies between accurate and approximate current densities evaluated using Eq.~\eqref{eq:discrepancy} for (a) $E_0 = 0.1$~V/{\AA} and (b) $E_0 = 1$~V/{\AA}.}
	\label{fig:2D_disrepancies}
\end{figure}

We quantify the discrepancy between an accurate current density, $J(t)$, and an approximate one, $J_N(t)$, using the following measure:
\begin{equation}\label{eq:discrepancy}
\delta = \frac{\max_t |J(t) - J_N(t)|}{\max_t |J(t)|}.
\end{equation}
Figure \ref{fig:2D_disrepancies} shows how $\delta$ depends on $\epsilon_{\mathrm{cut}}$.
For the moderately strong peak field equal to $E_0 = 0.1$~V/{\AA}, already the first-order correction decreases $\delta$ by two orders of magnitude in the range of cut-off energies between 25~eV and 200~eV.
The third-order correction decreases the discrepancy even further.
The improvement due to the analytical corrections is not as dramatic when the laser pulse has a peak field as large as $E_0 = 1$~V/{\AA}, but we see that the third-order correction is particularly useful in this case.
For example, without corrections, we get $\delta = 0.0022$ at $\epsilon_{\mathrm{cut}}=200$~eV; with the first-order corrections, this level of discrepancy is first reached at $\epsilon_{\mathrm{cut}}=176$~eV, while the third-order correction provides this value of $\delta$ already at $\epsilon_{\mathrm{cut}}=83$~eV.

\begin{figure}[!htb]
	\begin{tabular}{p{0.9\columnwidth} p{0pt}}
		\vspace{0pt} \includegraphics[width=\columnwidth]{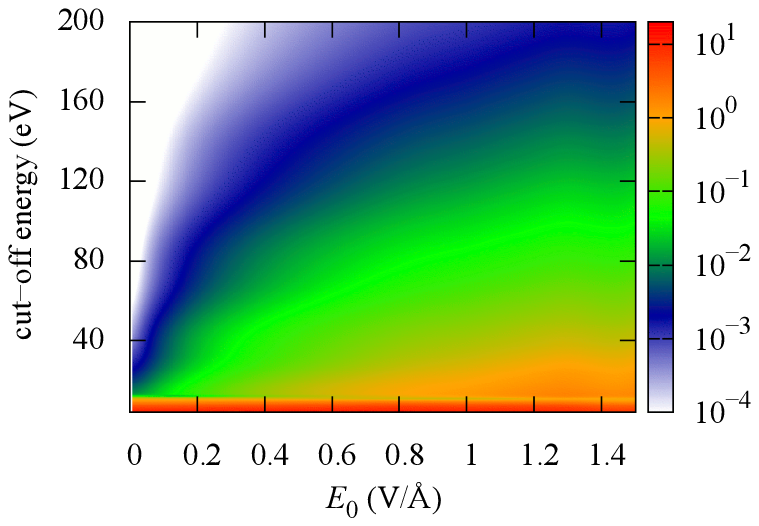} &
		\vspace{0pt} \hspace{-0.98\columnwidth}
		\textbf{(a)}
	\end{tabular}\\[-3mm]
	\begin{tabular}{p{0.9\columnwidth} p{0pt}}
		\vspace{0pt} \includegraphics[width=\columnwidth]{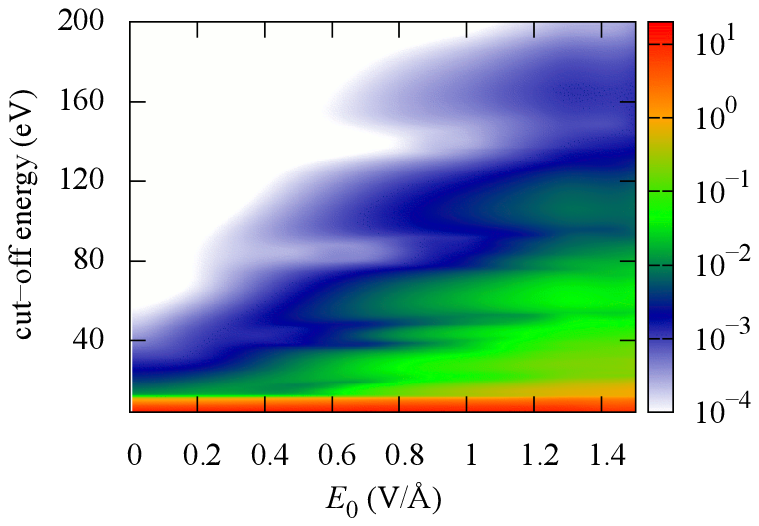} &
		\vspace{0pt} \hspace{-0.98\columnwidth}
		\textbf{(b)}
	\end{tabular}
  \caption{The discrepancy, $\delta$, between the accurate current density and that evaluated with corrections up to (a) the first and (b) the third order.}
  \label{fig:disrepancy_map}
\end{figure}

Figure \ref{fig:disrepancy_map} gives more details on how the discrepancy depends on the peak electric field.
In this figure, we represent $\delta$ using false colors.
Panel (a) shows the result of applying the first-order correction.
For panel (b), we corrected the current density using both $c_1$ and $c_3$.
This figure confirms the usefulness of the third-order correction.
In both panels, we observe the same general trend: a stronger laser field requires a larger value of $\epsilon_{\mathrm{cut}}$.
However, when the third-order correction is used, the dependence is not monotonous.
We also notice that the local minima of $\delta(\epsilon_{\mathrm{cut}})$ in Fig.~\ref{fig:disrepancy_map}(b) tend to maintain their positions as $E_0$ is increased.

\section{Conclusions and outlook}
We have shown how to significantly reduce the number of bands required by velocity-gauge simulations performed in a stationary basis of Bloch states.
The original motivation for this work was to speed up modeling of highly nonlinear ultrafast phenomena that take place when intense few-cycle laser pulses interact with crystalline solids.
However, our method is very general, and it may find other applications.
In particular, it can improve real-time methods of evaluating linear and nonlinear optical susceptibilities (the results presented here are only relevant to $\chi^{(1)}$, $\chi^{(2)}$, and $\chi^{(3)}$, but higher-order corrections can be derived as well).

Equations \eqref{eq:first-order_correction}, \eqref{eq:second-order_correction}, and \eqref{eq:third-order_correction} present the main results of this paper.
The key idea of our approach was to consider the limit where the frequency of the field, $\omega_0$, approaches zero, while the amplitude of the vector potential is fixed.
In this limit, the electric field becomes infinitesimally weak, which is a significant difference between our approach and that used to analyze the polarization divergence~\cite{Aspnes_PRB_1972,Sipe_PRB_1993}, where the low-frequency limit was applied with a fixed amplitude of the electric field, which made the vector potential grow indefinitely.

Our approach thrives on the fact that the adiabatic approximation works well for those contributions to the polarization response that involve highly excited stationary states, where the excitation energy is much larger than $\hbar\omega_0$.
Hence, our method works best when $\omega_0$ is small.
We have demonstrated excellent performance for a near-infrared laser pulse interacting with a model dielectric.

\section*{Acknowledgements}
The authors thank S.~Yu.~Kruchinin for his helpful remarks.
Supported by the DFG Cluster of Excellence: Munich-Centre for Advanced Photonics.
M.\,S.\,W.\ was supported by the International Max Planck Research School of Advanced Photon Science (IMPRS-APS).

\bibliographystyle{apsrev4-1}
\bibliography{yakovlev_personal,yakovlev_general,additional_references}

\end{document}